\documentclass[twoside,a4paper,final,onecolumn]{article}%
\usepackage{amssymb}
\usepackage{amsfonts}
\usepackage{amsmath}
\usepackage{graphicx}%
\providecommand{\U}[1]{\protect\rule{.1in}{.1in}}
\newtheorem{theorem}{Theorem}

\newtheorem{corollary}[theorem]{Corollary}

\newtheorem{lemma}[theorem]{Lemma}

\newtheorem{remark}[theorem]{Remark}

\newenvironment{proof}[1][Proof]{\noindent\textbf{#1.} }{\ \rule{0.5em}{0.5em}}
\begin{document}

\title{\textbf{Rotational and Self-similar Solutions for the Compressible Euler
Equations in }$R^{3}$}
\author{M\textsc{anwai Yuen\thanks{E-mail address: nevetsyuen@hotmail.com }}\\\textit{Department of Mathematics and Information Technology,}\\\textit{The Hong Kong Institute of Education,}\\\textit{10 Lo Ling Road, Tai Po, New Territories, Hong Kong}}
\date{Revised 24-Sept-2014}
\maketitle

\begin{abstract}
In this paper, we present rotational and self-similar solutions for the
compressible Euler equations in $R^{3}$ using the separation method. These
solutions partly complement Yuen's irrotational and elliptic solutions in
$R^{3}$ [\textit{Commun. Nonlinear Sci. Numer. Simul.} \textbf{17} (2012),
4524--4528] as well as rotational and radial solutions in $R^{2}$
[\textit{Commun. Nonlinear Sci. Numer. Simul.} \textbf{19 }(2014),
2172--2180]. A newly deduced Emden dynamical system is obtained. Some blowup
phenomena and global existences of the responding solutions can be determined.
The 3D rotational solutions provide concrete reference examples for vortices
in computational fluid dynamics.

\ 


\ 


\end{abstract}

\section{Introduction}

In fluid dynamics, the $N$-dimensional isentropic compressible Euler equations
are expressed as follows:%
\begin{equation}
\left\{
\begin{array}
[c]{rl}%
{\normalsize \rho}_{t}{\normalsize +\nabla\cdot(\rho\vec{u})} &
{\normalsize =}{\normalsize 0,}\\[0.08in]%
\rho\lbrack\vec{u}_{t}+(\vec{u}\cdot\nabla)\vec{u}]+K\nabla\rho^{\gamma} & =0,
\end{array}
\right.  \label{EulerEq}%
\end{equation}
where $\rho=\rho(t,\vec{x})$ denotes the density of the fluid, $\vec{u}%
=\vec{u}(t,\vec{x})=(u_{1},u_{2},\cdots,u_{N})\in R^{N}$ is the velocity,
$\vec{x}=(x_{1},x_{2},\cdots,x_{N})\in R^{N}$, that we use $x_{1}=x$,
$x_{2}=y$ and $x_{3}=z$ for $N\leq3$ and $K>0,\;\gamma\geq1$ are constants.

Basically, these Euler equations are a set of equations that govern the
inviscid flow of a fluid. The first and second equations of (\ref{EulerEq})
represent, respectively, the conservation of mass and the momentum of the fluid.

The Euler equations have applications in many mathematical physics subjects,
such as fluids, plasmas, condensed matter, astrophysics, oceanography and
atmospheric dynamics. For real-life applications, they can be used in the
study of turbulence, weather forecasting and the prediction of earthquakes and
the explosion of supernovas.

The Euler equations are the basic model of shallow water flows
\cite{ConstantinA}. In \cite{Einzel}, they are used to model the super-fluids
produced by Bose-Einstein condensates in the dilute gases of alkali metals, in
which identical gases do not interact at very low temperatures. However, at
the microscopic level, fluids or gases are formed by many tiny discrete
molecules or particles that collide with one another. As the cost of directly
calculating the particle-to-particle or molecule-to-molecule evolution of the
fluids on a large scale is expensive, approximation methods are needed to
considerably simplify the process. An example of an approximation method is
given in \cite{CIP}, where the Euler equations are used to describe the
behavior of fluids at the statistical limit of a large number of small ideal
molecules or particles by ignoring the less influential effects, such as
self-gravitational forces and the relativistic effect. The detailed derivation
of the Euler equations can be found in \cite{Lion} and \cite{CW}.

The construction of analytical or exact solutions is an important area in
mathematical physics and applied mathematics, as it can further classify
nonlinear phenomena. For non-rotational flows, Makino first obtained the
radial symmetry solutions for the Euler equations (\ref{EulerEq}) in $R^{N}$
in 1993 \cite{Makino93exactsolutions}. A number of special solutions for these
equations \cite{LW} and \cite{Yuen3DexactEuler} were subsequently obtained.
Yuen later obtained a class of self-similar solutions with elliptical symmetry
in 2012 \cite{YuenCNSNS2012}. For rotational flows, Zhang and Zheng
constructed explicitly rotational solutions for the Euler equations with
$\gamma=2$ and $N=2$ in 1997 \cite{ZZ}. In 2014, Yuen obtained a class of
rotational solutions for the compressible Euler equations (\ref{EulerEq}) for
$\gamma>1$ in 2D in \cite{YuenCNSNS2014}:%
\begin{equation}
\left\{
\begin{array}
[c]{c}%
\rho=\frac{\max\left(  \left(  -\frac{\lambda(\gamma-1)}{2K\gamma}\eta
+\alpha\right)  ^{\frac{1}{\gamma-1}},\text{ }0\right)  }{a^{2}(t)},\\
u_{1}=\frac{\dot{a}(t)}{a(t)}x-\frac{\xi}{a^{2}(t)}y\text{,}\\
u_{2}=\frac{\xi}{a^{2}(t)}x+\frac{\dot{a}(t)}{a(t)}y\text{,}\\
\ddot{a}(t)-\frac{\xi^{2}}{a^{3}(t)}=\frac{\lambda}{a^{2\gamma-1}(t)}\text{,
}a(0)=a_{0}>0\text{, }\dot{a}(0)=a_{1},
\end{array}
\right.  \label{2-Dg>1Rotation}%
\end{equation}
with a self-similar variable $\eta=\frac{x^{2}+y^{2}}{a^{2}(t)}$ and arbitrary
constants $\lambda$, $\alpha\geq0$, $\xi\neq0,$ $a_{0}$ and $a_{1}$.

For the physical applications of the similar solutions for the compressible
Euler equations, readers may refer to \cite{Sedov, Ba, Ba2, Barna, CN}.

Based on the works in \cite{YuenCNSNS2012} and \cite{YuenCNSNS2014}, we obtain
novel rotational and self-similar solutions for the 3D compressible Euler
equations (\ref{EulerEq}).

\begin{theorem}
\label{thm:1}For the compressible Euler equations (\ref{EulerEq}) in $R^{3}$,
there exists a family of rotational and self-similar solutions%
\begin{equation}
\left\{
\begin{array}
[c]{c}%
\rho=\frac{f\left(  s\right)  }{a^{2}(t)b(t)}\text{, }\\
u_{1}=\frac{\dot{a}(t)}{a(t)}x-\frac{\xi}{a^{2}(t)}y\text{,}\\
u_{2}=\frac{\xi}{a^{2}(t)}x+\frac{\dot{a}(t)}{a(t)}y\text{,}\\
u_{3}=\frac{\dot{b}(t)}{b(t)}z\text{,}%
\end{array}
\right.  \label{YuenSolution3DRotational}%
\end{equation}
with a variable $s=\frac{x^{2}+y^{2}}{a^{2}(t)}+\frac{z^{2}}{b^{2}(t)}$ and
\begin{equation}
f(s)=\left\{
\begin{array}
[c]{l}%
\alpha e^{-\frac{\lambda}{2K}s}\text{\quad\quad for }\gamma=1\text{,}%
\\[0.02in]%
\max\left(  \left(  -\frac{\lambda(\gamma-1)}{2K\gamma}s+\alpha\right)
^{\frac{1}{\gamma-1}},\text{ }0\right)  \text{ \ \ for }\gamma>1\text{,}%
\end{array}
\right.  \label{Solution5}%
\end{equation}
and the corresponding Emden system%
\begin{equation}
\left\{
\begin{array}
[c]{c}%
\ddot{a}(t)-\frac{\xi^{2}}{a^{3}(t)}=\frac{\lambda}{a^{2\gamma-1}%
(t)b^{\gamma-1}(t)},\text{ }a(0)=a_{0}>0,\text{ }\dot{a}(0)=a_{1}\text{,}\\
\ddot{b}(t)=\frac{\lambda}{a^{2\gamma-2}(t)b^{\gamma}(t)}\text{, }%
b(0)=b_{0}>0,\text{ }\dot{b}(0)=b_{1}\text{,}%
\end{array}
\right.  \label{Solution6}%
\end{equation}
where $\xi\neq0$, $\lambda$, $\alpha\geq0$, $a_{0}$, $a_{1}$, $b_{0}$ and
$b_{1}$ are arbitrary constants.\newline In particular, if any one following
condition is further fulfilled,\newline(1) with $\gamma=1$;\newline(2) with
$\gamma>1$,\newline(2a) $\lambda\leq0$ or\newline(2b) $\lambda>0$ and
$\gamma<2,$\newline solutions (\ref{YuenSolution3DRotational}%
)--(\ref{Solution6}) are $C^{1}$.
\end{theorem}

\begin{remark}
Solutions (\ref{YuenSolution3DRotational})--(\ref{Solution6}) of the
compressible Euler equations (\ref{EulerEq}) in $R^{3}$ are very efficient for
testing the accuracy of many numerical solutions about vortices in
computational fluid dynamics. In particular, the 3D rotational solutions
provide concrete reference examples for modeling typhoons in oceans.
\end{remark}

\begin{remark}
For the compressible Euler equations (\ref{EulerEq}) in $R^{3}$, the
rotational solutions (\ref{YuenSolution3DRotational})--(\ref{Solution6})
correspond to Yuen's irrotational and elliptic solutions in $R^{3}$
\cite{YuenCNSNS2012} as well as rotational and radial solutions in $R^{2}$
\cite{YuenCNSNS2014}.
\end{remark}

\section{ Rotational and Self-similar Solutions}

To prove Theorem 1, we need the following novel lemma for the
three-dimensional mass equation (\ref{EulerEq})$_{1}$.

\begin{lemma}
\label{lem:generalsolutionformasseqrotation3d}For the equation of the
conservation of mass (\ref{EulerEq})$_{1}$ in $R^{3}$,%
\begin{equation}
\rho_{t}+\nabla\cdot\left(  \rho\vec{u}\right)  =0, \label{MassLemma}%
\end{equation}
there exists a family of solutions,%
\begin{equation}
\left\{
\begin{array}
[c]{c}%
\rho=\frac{f(s)}{a^{2}(t)b(t)},\\
u_{1}=\frac{\dot{a}(t)}{a(t)}x-G(t)y,\\
u_{2}=G(t)x+\frac{\dot{a}(t)}{a(t)}y,\\
u_{3}=\frac{\dot{b}(t)}{b(t)}z,
\end{array}
\right.  \label{FunctionalLemma}%
\end{equation}
with a self-similar variable $s=\frac{x^{2}+y^{2}}{a^{2}(t)}+\frac{z^{2}%
}{b^{2}(t)}$ and arbitrary $C^{1}$ functions $f(s)\geq0$, $G(t)$, $a(t)>0$ and
$b(t)>0$.

\begin{proof}
By substituting the corresponding functions (\ref{FunctionalLemma}) for $\rho$
and $\vec{u}$ into the mass equation (\ref{MassLemma}) in $R^{3}$, we obtain%
\begin{equation}
\rho_{t}+\nabla\cdot\left(  \rho\vec{u}\right)
\end{equation}%
\begin{equation}
=\rho_{t}+\nabla\rho\cdot\vec{u}+\rho\nabla\cdot\vec{u}%
\end{equation}%
\begin{align}
&  =\frac{\partial}{\partial t}\left[  \frac{f(s)}{a^{2}(t)b(t)}\right]
+\frac{\partial}{\partial x}\left[  \frac{f(s)}{a^{2}(t)b(t)}\right]  \left(
\frac{\dot{a}(t)}{a(t)}x-G(t)y\right) \nonumber\\
&  +\frac{\partial}{\partial y}\left[  \frac{f(s)}{a^{2}(t)b(t)}\right]
\left(  G(t)x+\frac{\dot{a}(t)}{a(t)}y\right)  +\frac{\partial}{\partial
z}\left[  \frac{f(s)}{a^{2}(t)b(t)}\right]  \frac{\dot{b}(t)}{b(t)}%
z+\frac{f(s)}{a^{2}(t)b(t)}\left[  2\frac{\dot{a}(t)}{a(t)}+\frac{\dot{b}%
(t)}{b(t)}\right] \\
&  =-\frac{2\dot{a}(t)f(s)}{a^{3}(t)b(t)}-\frac{\dot{b}(t)f(s)}{a^{2}%
(t)b^{2}(t)}+\frac{\dot{f}(s)}{a^{2}(t)b(t)}\left[  \frac{x^{2}+y^{2}}%
{a^{3}(t)}\left(  -2\dot{a}(t)\right)  +\frac{z^{2}}{b^{3}(t)}(-2\dot
{b}(t))\right] \nonumber\\
&  +\frac{\dot{f}(s)}{a^{2}(t)b(t)}\frac{2x}{a^{2}(t)}\left(  \frac{\dot
{a}(t)}{a(t)}x-G(t)y\right)  +\frac{\dot{f}(s)}{a^{2}(t)b(t)}\frac{2y}%
{a^{2}(t)}\left(  G(t)x+\frac{\dot{a}(t)}{a(t)}y\right) \nonumber\\
&  +\frac{\dot{f}(s)}{a^{2}(t)b(t)}\frac{2z}{b^{2}(t)}\frac{\dot{b}(t)}%
{b(t)}z+\frac{f(s)}{a^{2}(t)b(t)}\left[  2\frac{\dot{a}(t)}{a(t)}+\frac
{\dot{b}(t)}{b(t)}\right] \\
&  =\frac{\dot{f}(s)}{a^{2}(t)b(t)}\left[  \frac{x^{2}+y^{2}}{a^{3}(t)}\left(
-2\dot{a}(t)\right)  +\frac{z^{2}}{b^{3}(t)}(-2\dot{b}(t))\right]  +\frac
{\dot{f}(s)}{a^{2}(t)b(t)}\frac{2x}{a^{2}(t)}\left(  \frac{\dot{a}(t)}%
{a(t)}x-G(t)y\right) \nonumber\\
&  +\frac{\dot{f}(s)}{a^{2}(t)b(t)}\frac{2y}{a^{2}(t)}\left(  G(t)x+\frac
{\dot{a}(t)}{a(t)}y\right)  +\frac{\dot{f}(s)}{a^{2}(t)b(t)}\frac{2z}%
{b^{2}(t)}\frac{\dot{b}(t)}{b(t)}z\\
&  =0\text{.}%
\end{align}
The proof is complete.
\end{proof}
\end{lemma}

We are now in a position to prove Theorem 1.

\begin{proof}
[Proof of Theorem 1]By the above lemma, functions
(\ref{YuenSolution3DRotational})--(\ref{Solution6}) can be applied to solve
the mass equation (\ref{EulerEq})$_{1}$ in $R^{3}$ with arbitrary $C^{1}$
functions $f(s)\geq0$, $a(t)>0$ and $b(t)>0$.\newline For the first momentum
equation (\ref{EulerEq})$_{2,1}$, we have%
\begin{align}
&  \rho\left(  u_{1t}+u_{1}u_{1x}+u_{2}u_{1y}+u_{3}u_{1z}\right)
+K\frac{\partial}{\partial x}\left[  \frac{f(s)}{a^{2}(t)b(t)}\right]
^{\gamma}\\
&  =\rho\left\{
\begin{array}
[c]{c}%
\frac{\partial}{\partial t}\left[  \frac{\dot{a}(t)}{a(t)}x-\frac{\xi}%
{a^{2}(t)}y\right]  +\left[  \frac{\dot{a}(t)}{a(t)}x-\frac{\xi}{a^{2}%
(t)}y\right]  \frac{\partial}{\partial x}\left[  \frac{\dot{a}(t)}%
{a(t)}x-\frac{\xi}{a^{2}(t)}y\right] \\
+\left[  \frac{\xi}{a^{2}(t)}x+\frac{\dot{a}(t)}{a(t)}y\right]  \frac
{\partial}{\partial y}\left[  \frac{\dot{a}(t)}{a(t)}x-\frac{\xi}{a^{2}%
(t)}y\right]
\end{array}
\right\} \nonumber\\
&  +K\gamma\frac{f^{\gamma-1}(s)}{a^{2\gamma}(t)b^{\gamma}(t)}\dot{f}%
(s)\frac{2x}{a^{2}(t)}\\
&  =\rho\left\{
\begin{array}
[c]{c}%
\left(  -\frac{\dot{a}^{2}(t)}{a^{2}(t)}+\frac{\ddot{a}(t)}{a(t)}\right)
x+2\frac{\xi\dot{a}(t)}{a^{3}(t)}y+\left[  \frac{\dot{a}(t)}{a(t)}x-\frac{\xi
}{a^{2}(t)}y\right]  \frac{\dot{a}(t)}{a(t)}\\
-\left[  \frac{\xi}{a^{2}(t)}x+\frac{\dot{a}(t)}{a(t)}y\right]  \frac{\xi
}{a^{2}(t)}%
\end{array}
\right\} \nonumber\\
&  +K\gamma\frac{f^{\gamma-1}(s)}{a^{2\gamma}(t)b^{\gamma}(t)}\dot{f}%
(s)\frac{2x}{a^{2}(t)}\\
&  =\rho\left\{  \left[  \frac{\ddot{a}(t)}{a(t)}-\frac{\xi^{2}}{a^{4}%
(t)}\right]  x+K\gamma\frac{f^{\gamma-2}(s)}{a^{2\gamma-2}(t)b^{\gamma-1}%
(t)}\dot{f}(s)\frac{2x}{a^{2}(t)}\right\} \\
&  =\frac{\rho}{a^{2\gamma-1}(t)b^{\gamma-1}(t)}\frac{x}{a(t)}\left\{  \left[
\left(  \ddot{a}(t)-\frac{\xi^{2}}{a^{3}(t)}\right)  a^{2\gamma-1}%
(t)b^{\gamma-1}(t)\right]  +2K\gamma f^{\gamma-2}(s)\dot{f}(s)\right\} \\
&  =\frac{\rho}{a^{2\gamma-1}(t)b^{\gamma-1}(t)}\frac{x}{a(t)}\left\{
\lambda+2K\gamma f^{\gamma-2}(s)\dot{f}(s)\right\} \\
&  =0,
\end{align}
where
\begin{equation}
\left\{
\begin{array}
[c]{c}%
\ddot{a}(t)-\frac{\xi^{2}}{a^{3}(t)}=\frac{\lambda}{a^{2\gamma-1}%
(t)b^{\gamma-1}(t)},\\
\text{ }a(0)=a_{0}>0,\text{ }\dot{a}(0)=a_{1}\text{,}%
\end{array}
\right.  \label{a(t)}%
\end{equation}
and
\begin{equation}
\left\{
\begin{array}
[c]{c}%
\lambda+2K\gamma f^{\gamma-2}(s)\dot{f}(s)=0,\\
f(0)=\alpha\geq0\text{.}%
\end{array}
\right.  \label{ODE}%
\end{equation}
The exact solution of ordinary differential equation (\ref{ODE}) is
\begin{equation}
f(s)=\left\{
\begin{array}
[c]{l}%
\alpha e^{-\frac{\lambda}{2K}s}\text{\quad\quad for }\gamma=1,\\[0.02in]%
\left(  -\frac{\lambda(\gamma-1)}{2K\gamma}s+\alpha\right)  ^{\frac{1}%
{\gamma-1}}\text{ \ \ \ for }\gamma>1.
\end{array}
\right.
\end{equation}
Therefore, to promise the non-negativeness of the $C^{1}$ density function
$\rho$, we can re-take $f(s)$ for $\gamma>1$ by a cut-off function%
\begin{equation}
f(s)=\max\left(  \left(  -\frac{\lambda(\gamma-1)}{2K\gamma}s+\alpha\right)
^{\frac{1}{\gamma-1}},\text{ }0\right)  \text{,}%
\end{equation}
choosing any one following additional condition,\newline(2a) $\lambda\leq0$
or\newline(2b) $\lambda>0$ and $\gamma<2$.\newline For the second momentum
equation (\ref{EulerEq})$_{2,2}$, we have%
\begin{align}
&  \rho\left(  u_{2t}+u_{1}u_{2x}+u_{2}u_{2y}+u_{3}u_{2z}\right)
+K\frac{\partial}{\partial y}\left[  \frac{f(s)}{a^{2}(t)b(t)}\right]
^{\gamma}\\
&  =\rho\left\{
\begin{array}
[c]{c}%
\frac{\partial}{\partial t}\left[  \frac{\xi}{a^{2}(t)}x+\frac{\dot{a}%
(t)}{a(t)}y\right]  +\left[  \frac{\dot{a}(t)}{a(t)}x-\frac{\xi}{a^{2}%
(t)}y\right]  \frac{\partial}{\partial x}\left[  \frac{\xi}{a^{2}(t)}%
x+\frac{\dot{a}(t)}{a(t)}y\right] \\
+\left[  \frac{\xi}{a^{2}(t)}x+\frac{\dot{a}(t)}{a(t)}y\right]  \frac
{\partial}{\partial y}\left[  \frac{\xi}{a^{2}(t)}x+\frac{\dot{a}(t)}%
{a(t)}y\right]
\end{array}
\right\} \nonumber\\
&  +K\gamma\frac{f^{\gamma-1}(s)}{a^{2\gamma}(t)b^{\gamma}(t)}\dot{f}%
(s)\frac{2y}{a^{2}(t)}\\
&  =\rho\left\{
\begin{array}
[c]{c}%
-2\frac{\xi\dot{a}(t)}{a^{3}(t)}x+\left(  -\frac{\dot{a}^{2}(t)}{a^{2}%
(t)}+\frac{\ddot{a}(t)}{a(t)}\right)  y\\
+\left[  \frac{\dot{a}(t)}{a(t)}x-\frac{\xi}{a^{2}(t)}y\right]  \frac{\xi
}{a^{2}(t)}+\left[  \frac{\xi}{a^{2}(t)}x+\frac{\dot{a}(t)}{a(t)}y\right]
\frac{\dot{a}(t)}{a(t)}%
\end{array}
\right\} \nonumber\\
&  +K\gamma\frac{f^{\gamma-1}(s)}{a^{2\gamma}(t)b^{\gamma}(t)}\dot{f}%
(s)\frac{2y}{a^{2}(t)}\\
&  =\rho\left\{  \left[  \frac{\ddot{a}(t)}{a(t)}-\frac{\xi^{2}}{a^{4}%
(t)}\right]  y+K\gamma\frac{f^{\gamma-2}(s)}{a^{2\gamma-2}(t)b^{\gamma-1}%
(t)}\dot{f}(s)\frac{2y}{a^{2}(t)}\right\} \\
&  =\frac{\rho}{a^{2\gamma-1}(t)b^{\gamma-1}(t)}\frac{y}{a(t)}\left\{  \left[
\left(  \ddot{a}(t)-\frac{\xi^{2}}{a^{3}(t)}\right)  a^{2\gamma-1}%
(t)b^{\gamma-1}(t)\right]  +2K\gamma f^{\gamma-2}(s)\dot{f}(s)\right\} \\
&  =\frac{\rho}{a^{2\gamma-1}(t)b^{\gamma-1}(t)}\frac{y}{a(t)}\left\{
\lambda+2K\gamma f^{\gamma-2}(s)\dot{f}(s)\right\} \\
&  =0.
\end{align}
For the third momentum equation (\ref{EulerEq})$_{2,3}$, we have%
\begin{align}
&  \rho\left(  u_{3t}+u_{1}u_{3x}+u_{2}u_{3y}+u_{3}u_{3z}\right)
+K\frac{\partial}{\partial z}\left[  \frac{f(s)}{a^{2}(t)b(t)}\right]
^{\gamma}\\
&  =\rho\left\{  \frac{\partial}{\partial t}\left[  \frac{\dot{b}(t)}%
{b(t)}z\right]  +\frac{\dot{b}^{2}(t)}{b^{2}(t)}z\right\}  +K\gamma
\frac{f^{\gamma-1}(s)}{a^{2\gamma}(t)b^{\gamma}(t)}\dot{f}(s)\frac{2z}%
{b^{2}(t)}\\
&  =\rho\frac{\ddot{b}(t)}{b(t)}z+K\gamma\frac{f^{\gamma-1}(s)}{a^{2\gamma
}(t)b^{\gamma}(t)}\dot{f}(s)\frac{2z}{b^{2}(t)}\\
&  =\rho\left(  \frac{\ddot{b}(t)}{b(t)}z+K\gamma\frac{f^{\gamma-2}%
(s)}{a^{2\gamma-2}(t)b^{\gamma+1}(t)}\dot{f}(s)2z\right) \\
&  =\frac{\rho z}{a^{2\gamma-2}(t)b^{\gamma+1}(t)}\left[  \ddot{b}%
(t)a^{2\gamma-2}(t)b^{\gamma}(t)+2K\gamma f^{\gamma-2}(s)\dot{f}(s)\right] \\
&  =\frac{\rho z}{a^{2\gamma-2}(t)b^{\gamma+1}(t)}\left[  \lambda+2K\gamma
f^{\gamma-2}(s)\dot{f}(s)\right] \\
&  =0,
\end{align}
where%
\begin{equation}
\left\{
\begin{array}
[c]{c}%
\ddot{b}(t)=\frac{\lambda}{a^{2\gamma-2}(t)b^{\gamma}(t)}\text{,}\\
b(0)=b_{0}>0,\text{ }\dot{b}(0)=b_{1}\text{.}%
\end{array}
\right.  \label{b(t)}%
\end{equation}
The local existence of solutions for the Emden system (\ref{Solution6}) can be
obtained by the fixed point theorem.

In addition, we can generally consider the corresponding weak solutions of the
Euler equations, which in the sense, the discontinuous points with measure
zero can be ignored. We can have the weak $C^{0}$ solutions
(\ref{YuenSolution3DRotational})--(\ref{Solution6}).\newline We complete the proof.
\end{proof}

The following corollary is a direct consequence of Theorem 1 by the standard
comparison theorem and the classical energy method of second order ordinary
differential equations (which readers may see Chapter 2 in \cite{Arnold} for
details). We note that the similar analysis for the Emden system
(\ref{Solution6}) has been shown by Lemma 7 in \cite{YuenJMAA2008a} and Lemma
3 in \cite{YuenCQG2009}.

\begin{corollary}
\label{Cor}For solutions (\ref{YuenSolution3DRotational})--(\ref{Solution6})
of the compressible Euler equations (\ref{EulerEq}) in $R^{3}$, we
have\newline(1) if $\lambda>0$, the solutions are global;\newline(2) if
$\lambda=0$ and\newline(2a) $b_{1}\geq0$, the solutions are global;\newline%
(2b) $b_{1}<0$, the solutions blow up in a finite time $T$;\newline(3) if
$\lambda<0$ and\newline(3a) $\gamma=1$, the solutions blow up in a finite time
$T;$\newline(3b) $\gamma>1$ and $b_{1}\leq0$, the solutions blow up in a
finite time $T$.
\end{corollary}

\begin{remark}
Solutions (\ref{YuenSolution3DRotational})--(\ref{Solution6}) can also solve
the following compressible Navier-Stokes equations in $R^{3}$,%
\begin{equation}
\left\{
\begin{array}
[c]{rl}%
{\normalsize \rho}_{t}{\normalsize +\nabla\cdot(\rho\vec{u})} &
{\normalsize =}{\normalsize 0}\text{,}\\[0.08in]%
\rho\lbrack\vec{u}_{t}+(\vec{u}\cdot\nabla)\vec{u}]+K\nabla\rho^{\gamma} &
=\mu\Delta\vec{u}\text{,}%
\end{array}
\right.
\end{equation}
with a constant $\mu>0$.
\end{remark}

\section{Conclusion and Discussion}

In this paper, we present a class of rotational and self-similar solutions for
the 3D compressible Euler equations using the separation method. These novel
solutions (\ref{YuenSolution3DRotational})--(\ref{Solution6}) partly
complement Yuen's irrotational and elliptic solutions in 3D
\cite{YuenCNSNS2012} as well as rotational and radial solutions in 2D
\cite{YuenCNSNS2014}. A newly deduced Emden dynamical system
\begin{equation}
\left\{
\begin{array}
[c]{c}%
\ddot{a}(t)-\frac{\xi^{2}}{a^{3}(t)}=\frac{\lambda}{a^{2\gamma-1}%
(t)b^{\gamma-1}(t)}\text{, }a(0)=a_{0}>0,\text{ }\dot{a}(0)=a_{1}\text{,}\\
\ddot{b}(t)=\frac{\lambda}{a^{2\gamma-2}(t)b^{\gamma}(t)}\text{, }%
b(0)=b_{0}>0,\text{ }\dot{b}(0)=b_{1}\text{,}%
\end{array}
\right.  \label{NewEmden3D}%
\end{equation}
is obtained.

We observe that some qualitative behavior of solutions (\ref{2-Dg>1Rotation})
in $R^{2}$ is significantly different from solutions
(\ref{YuenSolution3DRotational})--(\ref{Solution6}) in $R^{3}$. In particular,
by applying the classical energy method for the Emden equation
(\ref{2-Dg>1Rotation})$_{4}$, we can easily establish 2D time-periodic
solutions (\ref{2-Dg>1Rotation}) for $1\leq\gamma<2$ with $\lambda<0$. (See
Lemma 5 in \cite{YuenCNSNS2014}.). However, it is trivial to see that it is
not possible to have the 3D time-periodic solutions
(\ref{YuenSolution3DRotational})--(\ref{Solution6}) as function $\ddot
{b}(t)<0$.

The complementary case with $\gamma>1$, $\lambda<0$ and $b_{1}>0$ for
Corollary \ref{Cor} is unknown, as it is not easy to be determined by the
classical methods, the comparison theorem and the energy method. In future
research, the following problems are highly recommended to be
investigated.\newline1. Can we show the blowup or global existence for
solutions for the Emden system (\ref{NewEmden3D}) with $\gamma>1$, $\lambda<0$
and $b_{1}>0$?\newline2. Can we modify solutions
(\ref{YuenSolution3DRotational})--(\ref{Solution6}) to show the existence of
the corresponding $C^{1}$ solutions for $\lambda>0$ and $\gamma\geq2$?

\section{Acknowledgement}

The author thanks for the reviewers' valuable comments for improving the
quality of this paper. This work is partially supported by the Internal
Research Grant RG21/2013-2014R from the Hong Kong Institute of Education.

\end{document}